\documentclass[english,preprint,prb,aps]{revtex4}

\begin{document}

\title{Comment on ``Spin and cyclotron energies of electrons in GaAs/Ga$_{1-x}$Al$_{x}$As
quantum wells''}

\author{A.~A.~Kiselev}

\email{kiselev@eos.ncsu.edu}

\affiliation{North Carolina State University, Raleigh, North Carolina 27695-7911,
USA}

\author{E.~L.~Ivchenko}

\affiliation{A.~F.~Ioffe Physico-Technical Institute, 194021 St.~Petersburg,
Russia}

\begin{abstract}
In a recent publication, Pfeffer and Zawadzki {[}cond-mat/0607150;
Phys. Rev. B \textbf{74}, 115309 (2006)] attempted a calculation of
electron $g$ factor in III-V heterostructures. The authors emphasize
that their outcome is in strong discrepancy with our original result
{[}Ivchenko and Kiselev, Sov. Phys. Semicond. \textbf{26}, 827 (1992)]
and readily conclude that ``the previous theory of the $g$ factor
in heterostructures is inadequate''. We show here that the entire
discrepancy can be tracked down to an additional contribution missing
in the incomplete elimination procedure of Pfeffer and Zawadzki. This
mistake equally affects their ``exact'' and approximate results. When
the overlooked terms stemming from the nondiagonal Zeeman interaction
between light hole and spin-orbit-split valence states are taken into
account in the effective electron dispersion, the results of the both
approaches applied to the three-level $\mathbf{k}\cdot\mathbf{p}$
model become identical.
\end{abstract}
\maketitle
Recently Pfeffer and Zawadzki have published a paper on the electron
$g$ factor and effective mass in bulk GaAs-type semiconductors and
GaAs-based quantum well structures.\cite{Pfeffer} Particularly, in
Section~II.A, they considered the three-level model (the conventional
Kane model) and derived an equation for the electron $g$ factor,
$g^{*}$, in the conduction band of a bulk zinc-blende-lattice semiconductor,
keeping a contribution quadratic in the electron wave vector $\mathbf{k}$.
We calculated the similar contribution in 1992, the result is given
by Eqs.~(7)--(9) in Ref.~\onlinecite{Ivchenko92}. Figure~2 in
Ref.~\onlinecite{Pfeffer} shows the $g^{*}$ value as a function
of $k_{z}$ for the bulk conduction electron states with the wave
vector parallel to the magnetic field $\mathbf{B}\parallel z$. In
this figure, the dashed and dashed-dotted lines, calculated by using
the results of Refs.~\onlinecite{Pfeffer} and \onlinecite{Ivchenko92},
reveal a striking dicrepancy. Here we will show that the reason of
this discrepancy is an incomplete application of the elimination procedure
in Ref.~\onlinecite{Pfeffer} resulting in Eqs.~(11), (12), and
(17) where important contributions to effective electron dispersion
and electron $g$ factor are missing. As soon as these contributions
are taken into account, the procedure gives the result identical to
that obtained in Ref.~\onlinecite{Ivchenko92}.

Following Pfeffer and Zawadzki, we present closed equations for the
electron energy of the spin-up ($+$) and spin-down ($-$) states
of the $n$th Landau level as \begin{equation}
\mathcal{E}_{n}^{+}=F^{+}(n,k_{z},\mathcal{E}_{n}^{+}),\ {\cal E}_{n}^{-}=F^{-}(n,k_{z},{\cal E}_{n}^{-}),\label{eq:FF}\end{equation}
where the effective electron dispersion functions $F^{\pm}(n,k_{z},\mathcal{E})$
 are found by the elimination of the valence band envelopes in favor
of the conduction band envelope functions in the multiband eigenvalue
problem with the $\mathbf{k}\cdot\mathbf{p}$ Hamiltonian $H$ given
in Eq.~(8) of Ref.~\onlinecite{Pfeffer}. Collecting  only the terms
resulting from the off-diagonal components of the matrix $H$, \emph{directly}
coupling conduction electrons with individual valence states, one
can reproduce the result of Pfeffer and Zawadzki. In the three-level
model, taking into account the \textbf{$\mathbf{k}\cdot\mathbf{p}$}
interaction of the lowest conduction band $c\Gamma_{6}$ with the
upper valence bands $v\Gamma_{8}$, $v\Gamma_{7}$, and retaining
also \emph{all} free-electron terms,  this reads \begin{eqnarray}
F_{{\rm PZ}}^{+}(n,k_{z},\mathcal{E}) & = & \hbar\omega_{c}^{0}\left(n+\frac{1}{2}\right)\left[1-\frac{E_{P_{0}}}{3}\left(\frac{3}{2E_{1}^{0}}+\frac{1}{2E_{3}^{0}}+\frac{1}{G_{2}^{0}}\right)\right]\nonumber \\
 & + & \frac{1}{2}\mu_{B}B\left[g_{0}+\frac{2E_{P_{0}}}{3}\left(\frac{3}{2E_{1}^{0}}-\frac{1}{2E_{3}^{0}}-\frac{1}{G_{2}^{0}}\right)\right]\nonumber \\
 & + & \frac{\hbar^{2}k_{z}^{2}}{2m_{0}}\left[1-\frac{E_{P_{0}}}{3}\left(\frac{2}{E_{2}^{0}}+\frac{1}{G_{1}^{0}}\right)\right],\label{eq:F_plus}\end{eqnarray}
 \begin{eqnarray}
F_{{\rm PZ}}^{-}(n+1,k_{z},\mathcal{E}) & = & \hbar\omega_{c}^{0}\left(n+\frac{3}{2}\right)\left[1-\frac{E_{P_{0}}}{3}\left(\frac{3}{2E_{4}^{0}}+\frac{1}{2E_{2}^{0}}+\frac{1}{G_{1}^{0}}\right)\right]\nonumber \\
 & - & \frac{1}{2}\mu_{B}B\left[g_{0}+\frac{2E_{P_{0}}}{3}\left(\frac{3}{2E_{4}^{0}}-\frac{1}{2E_{2}^{0}}-\frac{1}{G_{1}^{0}}\right)\right]\nonumber \\
 & + & \frac{\hbar^{2}k_{z}^{2}}{2m_{0}}\left[1-\frac{E_{P_{0}}}{3}\left(\frac{2}{E_{3}^{0}}+\frac{1}{G_{2}^{0}}\right)\right],\label{eq:F_minus}\end{eqnarray}
 where \begin{eqnarray}
E_{1}^{0} & = & E_{0}+\frac{1}{2}\mu_{B}g_{0}B+\hbar\omega_{c}^{0}\left(n-\frac{1}{2}\right)+\frac{\hbar^{2}k_{z}^{2}}{2m_{0}}-\mathcal{E},\nonumber \\
E_{2}^{0} & = & E_{0}+\frac{1}{6}\mu_{B}g_{0}B+\hbar\omega_{c}^{0}\left(n+\frac{1}{2}\right)+\frac{\hbar^{2}k_{z}^{2}}{2m_{0}}-\mathcal{E},\nonumber \\
E_{3}^{0} & = & E_{0}-\frac{1}{6}\mu_{B}g_{0}B+\hbar\omega_{c}^{0}\left(n+\frac{3}{2}\right)+\frac{\hbar^{2}k_{z}^{2}}{2m_{0}}-\mathcal{E},\nonumber \\
E_{4}^{0} & = & E_{0}-\frac{1}{2}\mu_{B}g_{0}B+\hbar\omega_{c}^{0}\left(n+\frac{5}{2}\right)+\frac{\hbar^{2}k_{z}^{2}}{2m_{0}}-\mathcal{E},\nonumber \\
G_{1}^{0} & = & G_{0}-\frac{1}{6}\mu_{B}g_{0}B+\hbar\omega_{c}^{0}\left(n+\frac{1}{2}\right)+\frac{\hbar^{2}k_{z}^{2}}{2m_{0}}-\mathcal{E},\nonumber \\
G_{2}^{0} & = & G_{0}+\frac{1}{6}\mu_{B}g_{0}B+\hbar\omega_{c}^{0}\left(n+\frac{3}{2}\right)+\frac{\hbar^{2}k_{z}^{2}}{2m_{0}}-\mathcal{E}.\label{eq:EG}\end{eqnarray}
 Here, where possible, we use the notations of Ref.~\onlinecite{Pfeffer}:
$m_{0}$ is the free electron mass, $E_{g}$ is the band gap and $E_{0}\equiv-E_{g}$,
$\Delta$ is the spin-orbit splitting of valence band and $G_{0}\equiv-E_{g}-\Delta$,
$p_{cv}$ is the interband matrix element of the momentum operator
and $E_{P_{0}}=2p_{cv}^{2}/m_{0}$. As distinct from this reference,
we preserved $\hbar^{2}k_{z}^{2}/2m_{0}$ in the right-hand sides
of Eqs.~(\ref{eq:EG}).  Note that the system of Eqs.~(11)--(13),
as printed in Ref.~\onlinecite{Pfeffer}, contains minor inconsistencies
in respect to the Landau level index $n$ in definitions of $E_{2}^{0},E_{3}^{0},G_{1}^{0},G_{2}^{0}$,
entering both Eq.~(11) and (12), which, we believe, did not affect
actual calculations. Also, for convenience of comparison with Ref.~\onlinecite{Ivchenko92},
we left the free electron Lande factor $g_{0}$ in symbolic form (which,
of course, can be replaced with good accuracy by a value of 2).

Of course, the reader should be cautioned that accounting in the
elimination procedure only for direct couplings and neglecting
\emph{indirect} couplings can be a source of a major error, but
for the moment we will just follow Pfeffer and Zawadzki. Defining
the electron $g$ factor by \[
g^{*}=\lim_{B\to0}\frac{\mathcal{E}_{n}^{+}-\mathcal{E}_{n}^{-}}{\mu_{B}B}\]
and using Eqs.~(\ref{eq:F_plus}), (\ref{eq:F_minus}), and
(\ref{eq:EG}), one can evaluate $g^{*}$. In particular, in the
second order in $k_{z}$, $g^{*}$ can be presented as a sum
$g_{{\rm PZ}}+g'$, where \begin{equation} g_{{\rm
PZ}}=g_{0}^{*}+\frac{\hbar^{2}k_{z}^{2}}{2m_{0}}\
\frac{E_{P_{0}}}{3}\left[\frac{2m_{0}}{m_{0}^{*}}\left(\frac{1}{E_{0}^{2}}-\frac{1}{G_{0}^{2}}\right)-g_{0}^{*}\left(\frac{2}{E_{0}^{2}}+\frac{1}{G_{0}^{2}}\right)\right]\label{eq:gPZ}\end{equation}
with \begin{eqnarray} g_{0}^{*} & = &
g_{0}+\frac{2E_{P_{0}}}{3}\left(\frac{1}{E_{0}}-\frac{1}{G_{0}}\right)\label{eq:g_0star}\end{eqnarray}
 is the result of Pfeffer and Zawadzki and \begin{equation}
g'=\frac{\hbar^{2}k_{z}^{2}}{2m_{0}}\left[-\frac{2E_{P_{0}}}{3}\left(\frac{1}{E_{0}^{2}}-\frac{1}{G_{0}^{2}}\right)+\frac{g_{0}E_{P_{0}}}{9}\left(\frac{2}{E_{0}^{2}}-\frac{1}{G_{0}^{2}}\right)\right].\label{eq:g_prime}\end{equation}
The small contribution $g'$ is absent in Ref.~\onlinecite{Pfeffer}
because in Eqs.~(\ref{eq:EG}), first, the free electron terms $\hbar^{2}k_{z}^{2}/2m_{0}$,
and later also those proportional to $\mu_{B}g_{0}B$ are disregarded.

Now, it is important to stress that neither $g_{{\rm PZ}}$ nor $g_{{\rm PZ}}+g'$
 can be correct in the three-level model because the both \emph{do
not pass the important test}, namely, they do not reduce to $g_{0}$
 when the spin-orbit splitting of the valence band is set to zero.

The negative test means that an additional contribution to effective
electron dispersions $F^{\pm}$ and $g^{*}$ should be sought. And,
indeed, such a contribution does exist, it stems from the indirect
coupling associated with the nondiagonal Zeeman interaction between
valence states, and equally affects ``exact'' Eqs.~(11), (12), ``exact''
result for electron $g$ factor, and its quadratic-in-$k$ expansion.
This contribution can be calculated analytically exactly (somewhat
lengthy), or, alternatively, captured in the iterative elimination
procedure if the terms cubic in the off-diagonal components of the
Hamiltonian $H$ are included in the consideration. The sought terms,
with sufficient accuracy, are \begin{eqnarray}
F_{(3)}^{+}(n,k_{z},\mathcal{E}) & \approx & \frac{H_{11,5}H_{5,7}H_{7,11}+H_{11,7}H_{7,5}H_{5,11}}{E_{0}G_{0}}\nonumber \\
 & = & \frac{2}{E_{0}G_{0}}\left(\sqrt{\frac{2}{3}}\frac{\hbar}{m_{0}}p_{cv}k_{z}\right)\left(\frac{\sqrt{8}}{3}\frac{\mu_{B}g_{0}B}{2}\right)\left(\frac{1}{\sqrt{3}}\frac{\hbar}{m_{0}}p_{cv}k_{z}\right)\nonumber \\
 & = & \frac{1}{2}\mu_{B}B\times\ \frac{\hbar^{2}k_{z}^{2}}{2m_{0}}\ \frac{8g_{0}E_{P_{0}}}{9E_{0}G_{0}}\label{eq:F3_plus}\end{eqnarray}
 and \begin{eqnarray}
F_{(3)}^{-}(n,k_{z},\mathcal{E}) & \approx & \frac{H_{4,12}H_{12,14}H_{14,4}+H_{4,14}H_{14,12}H_{12,4}}{E_{0}G_{0}}\nonumber \\
 & = & -\frac{1}{2}\mu_{B}B\ \times\frac{\hbar^{2}k_{z}^{2}}{2m_{0}}\ \frac{8g_{0}E_{P_{0}}}{9E_{0}G_{0}}=-F_{(3)}^{+}(n,k_{z},\mathcal{E}).\label{eq:F3_minus}\end{eqnarray}
 The Bloch states $c\Gamma_{6},v\Gamma_{8},v\Gamma_{7}$ are enumerated
from 4 to 7 and from 11 to 14 in accordance with Ref.~\onlinecite{Pfeffer}.
In particular, $H_{5,7}$ is the matrix element between the light-hole
and spin-orbit-split states with the angular-momentum component $+1/2$.
It follows then that the total electron $g$ factor is given by the
sum \begin{equation}
g^{*}=g_{{\rm PZ}}+g'+g^{(3)},\label{eq:total}\end{equation}
 where \begin{equation}
g^{(3)}=\frac{\hbar^{2}k_{z}^{2}}{2m_{0}}\ \frac{8g_{0}E_{P_{0}}}{9E_{0}G_{0}}.\label{eq:g3}\end{equation}
 One can check that now, at zero spin-orbit splitting of the valence
band, the value of $g^{*}$ reduces to $g_{0}$ as required. By using
Eqs.~(\ref{eq:gPZ}), (\ref{eq:g_prime}), (\ref{eq:total}), and
(\ref{eq:g3}) one can straightforwardly come to \[
g^{*}(k_{z})=g_{0}^{*}+(h_{1}+h_{2})k_{z}^{2},\]
 where  \[
h_{1}=h_{2}+\frac{4}{9}\frac{\hbar^{2}p_{cv}^{4}}{m_{0}^{3}}\frac{\Delta}{E_{g}(E_{g}+\Delta)}\left[\frac{4}{E_{g}^{2}}+\frac{2}{(E_{g}+\Delta)^{2}}+\frac{3}{E_{g}(E_{g}+\Delta)}\right],\]
 \[
h_{2}=-\frac{2}{9}g_{0}\frac{\hbar^{2}p_{cv}^{2}}{m_{0}^{2}}\frac{\Delta^{2}}{E_{g}^{2}(E_{g}+\Delta)^{2}}.\]
The above equation exactly coincides with the expansion presented
by Ivchenko and Kiselev in Ref.~\onlinecite{Ivchenko92} for the
Kane model in the magnetic field $\mathbf{B}$ parallel to $\mathbf{k}$.

In conclusion, the adequate application of the approaches applied
in Refs.~\onlinecite{Pfeffer} and \onlinecite{Ivchenko92} produces
identical results. That is true both for the exact result and quadratic-in-$k$
expansion. A few words are due about the remote band contributions
and their appropriate treatment in the $g$ factor theory. In case
of III-V materials, two terms, $C$ and $2C'$ define influence of
the remote bands on the effective mass and $g$ factor, respectively,
at the bottom of the conduction band. They can be derived in the second
order of the \textbf{$\mathbf{k}\cdot\mathbf{p}$} perturbation theory.
Thus, accounting for $2C'$ term in Eq.~(\ref{eq:g_0star}) for $g_{0}^{*}$,
calculated in the same order, is completely straightforward. At the
same time, preserving $C$ and $2C'$ in higher (fourth) order terms,
proportional to $k^{2}$, is typically a false accuracy and should
be avoided.

\end{document}